\documentclass[twocolumn,showpacs,amsmath,amssymb,aps]{revtex4}
\usepackage[dvipdfm]{graphicx}
\usepackage{rotating}
\usepackage{subfigure}
\usepackage{amsmath}
\usepackage{amsfonts}
\usepackage{amssymb}

\begin{document}

\title{Wang-Landau sampling in three-dimensional polymers}
\author{A. G. Cunha Netto\footnote{Corresponding author:\\
cunha@posgrad.ufg.br}}
\affiliation{Instituto de F\'{\i}sica, Universidade Federal de
Goi\'{a}s, C.P.131, 74.001-970,
Goi\^{a}nia (GO), Brazil}
\author{C. J. Silva}
\affiliation{Instituto de F\'{\i}sica, Universidade Federal de
Goi\'{a}s, C.P.131, 74.001-970,
Goi\^{a}nia (GO), Brazil}
\author{A. A. Caparica\footnote{caparica@if.ufg.br}}
\affiliation{Instituto de F\'{\i}sica, Universidade Federal de
Goi\'{a}s, C.P.131, 74.001-970,
Goi\^{a}nia (GO), Brazil}
\author{R. Dickman}
\affiliation{Departamento de F\'{\i}sica, Instituto de Ci\^{e}ncias Exatas,
Universidade Federal de Minas Gerais, C.P.702, Belo Horizonte (MG), Brazil}

\begin{abstract}
Monte Carlo simulations using Wang-Landau sampling are performed to
study three-dimensional chains of homopolymers on a lattice. We
confirm the accuracy of the method by calculating the thermodynamic
properties of this system. Our results are in good agreement with
those obtained using Metropolis importance sampling. This algorithm
enables one to accurately simulate the usually hardly accessible
low-temperature regions since it determines the density of states in
a single simulation.
\end{abstract}


\maketitle

\section{Introduction}

With the rapid development of computer processors, numerical
simulations using the Monte Carlo method have become a well
established tool for the study of proteins, polymers and spin-glass
models. When a Monte Carlo simulation using Metropolis importance
sampling \cite{metropolis} is carried out at a fixed temperature,
the quality of the data depends on the distance from criticality,
and therefore, multiple computer runs should be performed for each
temperature. In order to speed up the simulation several methods
have been suggested to overcome problems such as slow dynamics which
makes the Metropolis algorithm inefficient, and to study systems
with a rough energy landscape with multiple local minima in free
energy.  Important examples are the cluster-flip Swendsen-Wang
algorithm \cite{swang}, which has been used to reduce critical
slowing down at second order phase transitions, the multicanonical
method \cite{berg} which was introduced to overcome the barrier
between coexisting phases at first order phase transitions, the
broad histogram method \cite{oliveira} which directly calculates the
density of states with only one computer run, and the flat histogram
method \cite{jswang} which generates a flat histogram in energy
space similar to multicanonical simulations. The drawback of the
flat histogram method is the slow diffusion of the random walk which
is the same as in the multicanonical method. Nevertheless, 
no method is more efficient than that recently proposed by Wang and Landau
\cite{landau,landaupre,landaubjp,landauam} which allows one to get around these difficulties even
for large systems.

In this work we present results of simulations of three-dimensional
lattice polymer chains \cite{gennes} using Wang-Landau sampling, and
calculate thermodynamic properties of the system. The method is described 
in Sec. II.
Sec. III provides a brief background on the algorithm.
The results are given in Sec. IV and we summarize and conclude in Sec. V.

\section{The Model}

\begin{figure} [!htb]
\centering
\includegraphics[scale=0.57]{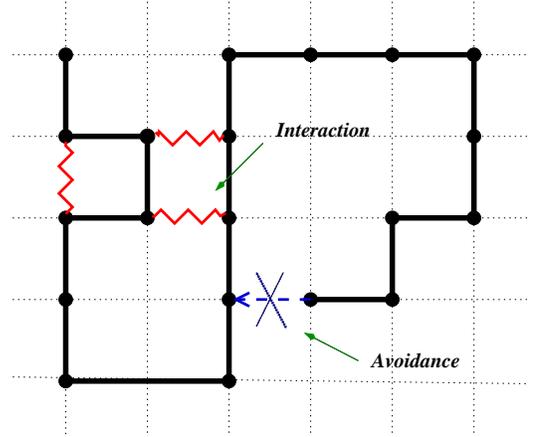}
\caption{\textit{Interaction Self Avoiding Walk.}}
\label{inter}
\end{figure}

We consider a homopolymer consisting of $N$ monomers which may
assume any self avoiding walk (SAW) configuration on a
three-dimensional lattice. Assuming that the polymer is in a bad
solvent, there is an effective monomer-monomer attraction in
addition to the self-avoidance constraint representing excluded
volume. For every pair of nonbonded nearest-neighbor monomers the
energy of the polymer is reduced by $\varepsilon$ (See Fig.
\ref{inter}).

Therefore, the interaction between two nonbonded monomers
$i$ and $j$ is given by
\begin{equation}
u_{i,j}=\left\{\begin{array}{c}\infty  \qquad \ \mbox r<\delta \\
-\varepsilon  \qquad \mbox r=\delta \\
0   \qquad \ \mbox r>\delta \\
\end{array}
\right.
\label{potential}
\end{equation}
where $\delta$ is the lattice constant.

The Hamiltonian for the model can be written as
\begin{equation}
\mathcal{H}=-\varepsilon\sum_{<i,j>}\sigma_{i}\sigma_{j}
\label{hamiltoniana}
\end{equation}
where $\sigma=1$ $(0)$ if the site $i$ is occupied (vacant), and the sum is
over nearest-neighbor pairs \cite{dickman}. (The sum is understood to exclude
pairs of bonded segments along the chain). We are interested in the equilibrium
properties at temperature $T$. (In the following, temperature is measured
in units of $\varepsilon/k_{B}$, where $k_{B}$ is the Boltzman's constant.)

In order to generate a Markov process for sampling configurations we have
first to define a protocol of movements. We used the so-called reptation
or \textit{``slithering snake''} move which consists of randomly adding a 
monomer to one end of the chain and removing a monomer from the other end,
maintining the size of the polymer constant.
We define
one Monte Carlo step as $N$ attempted moves. At certain moments, one end of
the chain may not be able to move, but successive motions of the other 
end release it from the trap. 

\section{The Method}

The Wang-Landau algorithm calculates the density of states $g(E)$ by
carrying out a random walk in energy space with an acceptance
probability proportional to $1/g(E)$ instead of the usual Boltzmann
weight $e^{-E/k_{B}T}$ used in the conventional Monte Carlo
simulation. The probability of energy $E$ changes therefore from
$g(E)e^{-E/k_{B}T}$ in the Metropolis scheme to an almost constant
probability in the Wang-Landau method. As a result the probability
minimum which appears at first order phase transitions, for example,
practically vanishes. The simulation is performed such that if
$E_{1}$ is the energy of the current configuration and $E_{2}$ is
the energy of a possible new configuration, the acceptance
probability is given by
\begin{equation}
p(E_{1}\rightarrow E_{2})=min\left[ \frac{g(E_{1})}{g(E_{2})},1 \right].
\end{equation}
For each accepted configuration we accumulate an energy histogram
$H(E)$. Since the density of states is not known \textit{a priori},
Wang and Landau proposed to set $g(E) = 1$ initially, for all energy
levels.

To study large systems the energy space is divided into bins and the
random walk is carried out independently in each segment. As the
random walk in energy space is performed, whenever a move to a
configuration with energy $E$ is accepted the density of states is
updated by multiplying it by a ``modification factor" $f>1$ that
accelerates the diffusion of the random walk, and an unit is added
to the histogram $H(E)$, i.e.
\begin{equation}
g(E) \rightarrow g(E) \cdot f;  ~~~~ \rm{and} ~~~~ \it{H(E) \rightarrow H(E)}\rm{+1}.
\end{equation}
The initial choice of $f$ is $f_{0}=e=2.71828...$.
The density of states is multiplied by $f$ until the accumulated histogram $H(E)$ is flat.
We then reduce $f$ by setting
\begin{equation}
f \rightarrow \sqrt{f},
\end{equation}
and resetting $H(E)=0$ for all energy values.

This process is repeated; the simulation converges to the true value
of $g(E)$ when $f$ is approximately $1$. In our simulations the
criterion of flatness was taken as each value of the histogram
reaching at least $80\%$ of the mean value $\langle H(E) \rangle$.
The histograms are generally checked after each $1000$ Monte Carlo
steps.

\section{Results}

\begin{figure}[!t]
\centering
\includegraphics[scale=0.65]{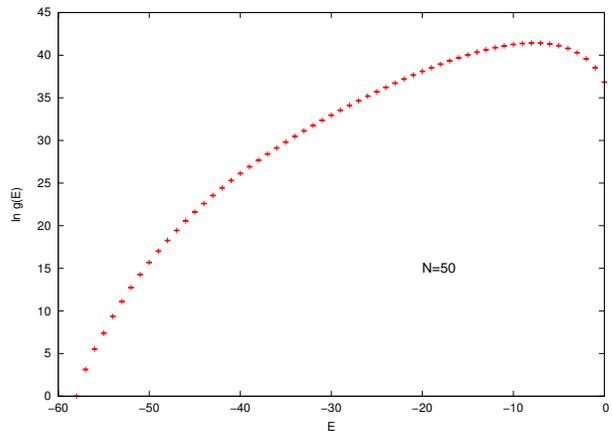}
\caption{Density of states for a chain of $N=50$ monomers. The ground state of this chain is $E_{g}=-58$.}
\label{g50}
\end{figure}

In this section we present results obtained for chains of lattice
homopolymers using the Wang-Landau algorithm and compare them with
those obtained using Metropolis algorithm importance sampling. In
Fig. \ref{g50} we show the density of states for a chain of $50$
monomers.

\begin{figure}[!h]
\centering
\includegraphics[scale=0.65]{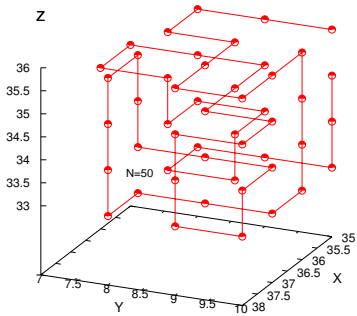}
\caption{A ground state configuration for $N=50$. In this case the energy
is $E_{g}=-58$.}
\label{caracol}
\end{figure}

The ground state is achieved by shaping a configuration rolled up
 like a snail (see Fig. \ref{caracol}). The energy of this
configuration is obtained by counting the number of nonbonded
nearest-neighbor pairs. In our simulations the polymer was always
set initially in this configuration and, using the algorithm
described above, the density of states was estimated as a result of
a random walk in the energy space.

Knowing the density of states, one can calculate any thermodynamic property
$A$ of the system through the canonical average
\begin{equation}
\langle A \rangle_{T} = \frac{\sum_{E} \langle A \rangle_{E} g(E)e^{-E/k_{B}T}}{Z}
\label{mean_of_A}
\end{equation}
where $\langle A \rangle_{E}$ is the microcanonical average of
observable $A$ obtained during the simulation, $g(E)$ is the density
of states for each energy level $E$, and $Z$ is the partition
function
\begin{equation}
Z = \sum_{E}g(E)e^{-E/k_{B}T}.
\end{equation}

In Monte Carlo simulations using Metropolis sampling we fix the
temperature and determine the thermodynamic average $\langle A
\rangle_{T}$. Therefore a new simulation is needed for each value of
temperature. In Wang-Landau sampling we estimate the density of
states and then calculate any thermodynamic property of interest by
means of a simple algebraic operation. The computational time in our
simulations was about ten times smaller using the Wang-Landau method
as compared with the Metropolis algorithm.

\begin{figure}[!t]
\centering
\includegraphics[scale=0.6]{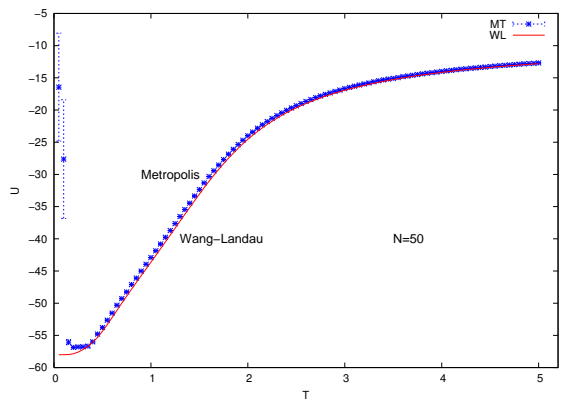}
\caption{Temperature dependence of the internal energy for a chain of $N=50$ monomers.}
\label{EWLMT}
\end{figure}

In Fig. \ref{EWLMT} we show the result for the internal energy via
Wang-Landau sampling, evaluated using the relation
\begin{equation}
U(T)=\dfrac{\sum_{E}Eg(E)e^{-E/k_{B}T}}{\sum_{E}g(E)e^{-E/k_{B}T}}\equiv \langle E \rangle,
\end{equation}
and compare it with that obtained via the Metropolis algorithm.

The specific heat can be determined from the fluctuations in the internal energy
\begin{equation}
C_{v}(T)=\frac{\left\langle E^{2}\right\rangle_{T}- \left\langle E\right\rangle^{2}_{T}}{k_{B}T^{2}}.
\label{specific_heat}
\end{equation}

\begin{figure}[!t]
\centering
\includegraphics[scale=0.6]{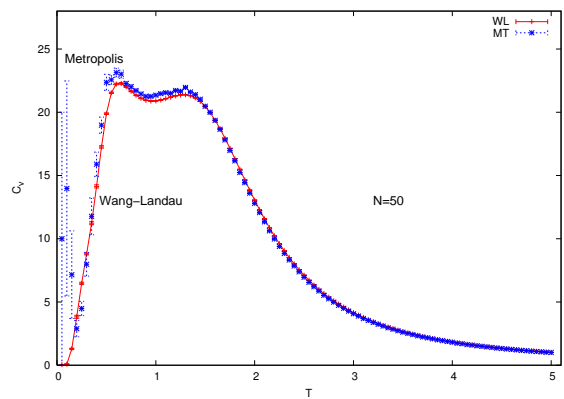}
\caption{Specific heat {\it vs.} temperature for a chain of $N=50$ monomers.}
\label{CvWLMT}
\end{figure}

The temperature dependence of the specific heat is shown in Fig.
\ref{CvWLMT}. Much more CPU time would be needed to obtain
significantly better results using the Metropolis algorithm. One can
see clearly from these results that the Wang-Landau method yields a
better description in the low-temperature regime.

\begin{figure}[!t]
\centering
\includegraphics[scale=0.6]{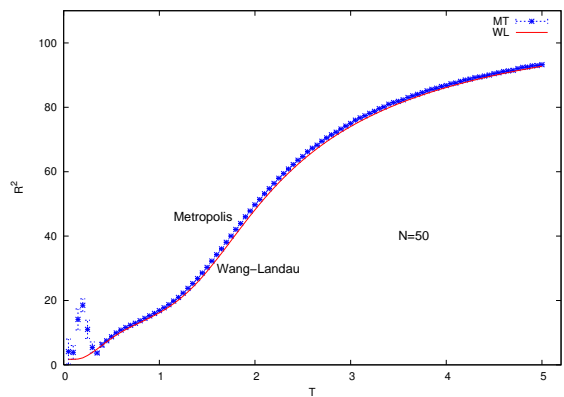}
\caption{Mean square end-to-end distance {\it vs.} temperature for a
chain of $N=50$ monomers.}
\label{R2WLMT}
\end{figure}

The temperature dependence of the mean square end-to-end distance,
given by
\begin{equation}
\left\langle R^{2}\right\rangle =\left\langle [(x_{N}-x_{0})^{2}+(y_{N}-y_{0})^{2}+(z_{N}-z_{0})^{2}]\right\rangle 
\label{R2}
\end {equation}
is shown in Fig. \ref{R2WLMT}.

Differentiating expression \ref{mean_of_A} with regard to temperature using
$\left\langle R^{2}\right\rangle $ as the observable $A$, we obtain the fluctuation of the
mean square end-to-end distance as
\begin{equation}
\frac{d\left\langle R^{2}\right\rangle_{T} }{dT}=\frac{\left\langle ER^{2}\right\rangle_{T}- \left\langle E\right\rangle_{T}\left\langle R^{2}\right\rangle_{T}}{T^{2}}.
\label{dR2}
\end{equation}

\begin{figure}[!t]
\centering
\includegraphics[scale=0.6]{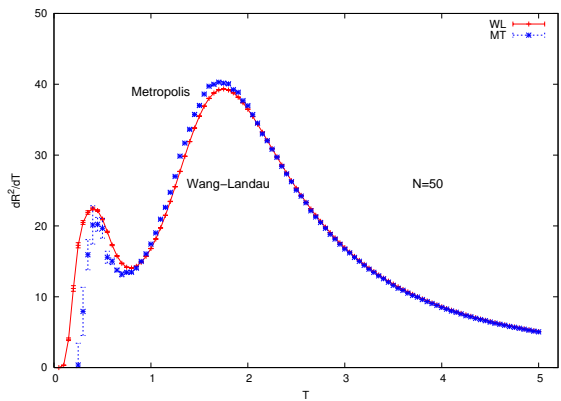}
\caption{Derivative of the mean square end-to-end distance {\it vs.} temperature. The curve with a peak resembles a second-order phase transition.}
\label{DR2WLMT}
\end{figure}

The temperature dependence of the derivative of the mean square end-to-end distance is shown
in Fig. \ref{DR2WLMT}. The coil-globule phase transition is characterized by a
peak in the derivative of $\left\langle R^{2}\right\rangle $ similar to the specific heat. 
We believe that
the low-temperature peak corresponds to excitations of the surface of the typlically
compact, folded configuration.  This effect  
is more pronounced in short chains.

An advantage of this algorithm is that we can readily calculate the
free energy and entropy, quantities that are not directly accessible
in conventional Monte Carlo simulations. In terms of the density of
states, the free energy can be calculated from the partition
function
\begin{equation}
F(T)=-k_{B}T\ln(Z)=-k_{B}T\ln\left(\sum_{E}g(E)e^{-E/k_{B}T}\right).
\end{equation}
\begin{figure}[!t]
\centering
\includegraphics[scale=0.6]{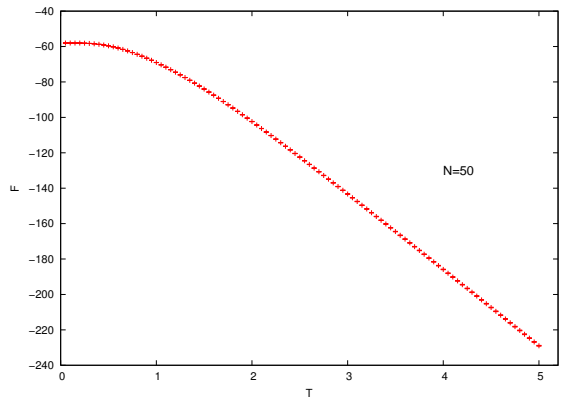}
\caption{Free energy {\it vs.} temperature for a chain of $N=50$ monomers.}
\label{free_energy}
\end{figure}

The temperature dependence of the free energy obtained from our simulation is shown in
Fig. \ref{free_energy} for a homopolymer with $N=50$ monomers.

The entropy is a key thermodynamic quantity that cannot be calculated
directly by conventional Monte Carlo simulation. It can be estimated by integration
over other thermodynamic quantities, such as specific heat, but these calculations are
somewhat unreliable since the specific heat itself is not easy to estimate accurately.
With an accurate density of states estimated by the Wang-Landau method, the entropy can be calculated simply using
\begin{equation}
S(T)=\frac{E(T)-F(T)}{T}.
\end{equation}

\begin{figure}[!h]
\centering
\includegraphics[scale=0.6]{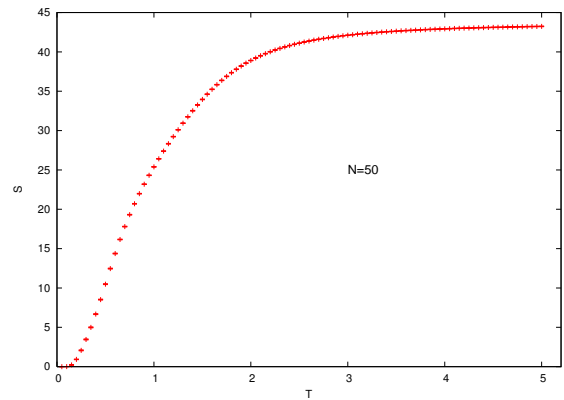}
\caption{Entropy {\it vs.} temperature for a chain of $N=50$ monomers.}
\label{entropy}
\end{figure}

In Fig. \ref{entropy} we show our result for the temperature dependence of the
entropy for a homopolymer chain of $N=50$ monomers.

\section{Conclusions}

We study the thermodynamic behavior of a three-dimensional
homopolymer chain in lattice using the Wang-Landau sampling. We
show that the density of states directly obtained by this algorithm
enables one to calculate thermodynamic properties even for large
systems with only one computer run. We also note from the
temperature dependence of the specific heat and the mean square
end-to-end distance that the low-temperature region is better
explored by this method than by standard Monte Carlo simulation.
Finally, thermodynamic quantities such as the free energy and the
entropy, which are not easily accessible by conventional methods,
are directly estimated from the density of states.

\begin{acknowledgments}
The authors would like to thank the Coordena\c{c}\~{a}o de
Aperfei\c{c}oamento de Pessoal de N\'{i}vel Superior (CAPES), the Conselho Nacional de
Desenvolvimento Cient\'{i}fico e Tecnol\'{o}gico (CNPq) and the Funda\c{c}\~{a}o de
Apoio \`{a} Pesquisa (FUNAPE-UFG) for financial support.
\end{acknowledgments}

\end{document}